\documentclass[12pt]{article}
\def\N{\cal{N}}
\def\Kahler{K{\rm{\ddot{a}}}hler}
\def\be{\begin{equation}}
\def\ee{\end{equation}}

\begin{document}
\newcommand{\mpl}{M_{\mathrm{Pl}}}
\setlength{\baselineskip}{18pt}
\begin{titlepage}
\begin{flushright}
KOBE-TH-01-01 \\
\end{flushright}

\vspace{2.5cm}

\centerline{{\LARGE\bf Constraints on Non-Holomorphic Correction}}
\vspace{5mm}
\centerline{{\LARGE\bf in ${\cal{N}}$=2 Superspace}}
\vspace{1cm}

\centerline{ Hiroyuki Yamashita \footnote{e-mail : 
hy257@phys.sci.kobe-ac.jp}
}
\vspace{1cm}
\centerline{{\it Graduate School of Science and Technology, 
Kobe University,}}
\centerline{\it Rokkodai, Nada, Kobe 657-8501, Japan}
%

%
%
%
\vspace{2cm}
\centerline{\large\bf Abstract}
\vspace{0.5cm}
We study quantum corrections on four derivative term of vector
multiplets in ${\cal{N}}=2$ supersymmetric Yang-Mills
theory. We first show splitting of quantum correction on gauge
neutral hypermultiplets from  $U(1)$ vector multiplets at four
derivative order. We then revisit the non-renormalization theorem
given by N. Seiberg and M. Dine and show the non-renormalization
theorem in mixed (Coulomb plus Higgs) branch even though gauge neutral
hypermultiplet develops the vacuum expectation value.
\end{titlepage}

%
%

\section{Introduction}
In the last decade supersymmetric field theories have been much
explored and exact solutions are well-often obtained not only
perturbatively but also non-perturbatively. The key ingredient
is $holomorphic$ argument ( See \cite{seiberg} for reviews of
${\cal{N}}=1$ supersymmetric field theories). For ${\cal{N}}=2$
SYM theory in four dimensions, the low energy effective action
at leading order is completely determined by ${\cal{N}}=2$
superspace chiral integral of holomorphic function \cite{SW}.
Notice that K${\rm{\ddot{a}}}$hler
potential is written as non-holomorphic term from ${\cal{N}}=1$
supersymmetric point of view. More supersymmetry, more controls of
quantum corrections! For ${\cal{N}}=4$ SYM theory we expect it is
strongly constrained. However, analysis is hampered because of the
lack of manifest ${\cal{N}}=4$ superspace. If such a formalism is
known, we would directly obtain the non-renormalization theorem of
four-derivative term, which is expected to be chiral integral in
${\cal{N}}=4$ superspace. Despite this obstacle, M. Dine and
N. Seiberg have shown the non-renormalization theorem
\cite{non-renormalization theorem} of four derivative operators of
purely chiral multiplets by taking advantage of ${\cal{N}}=2$
superspace  formulation and symmetry arguments in Coulomb branch
of $SU(2)$ gauge theory without turning on the vacuum expectation
value of gauge neutral
hypermultiplet. This result has been extended to $SU(N)$ gauge theory
in several papers \cite{extension, 2n derivative, study} and these
results are important in the context of recent studies of Matrix
theory, and in the ADS/CFT correspondence of string theory.
Although extra symmetries are assigned appropriately to be 
equivalent to ${\cal{N}}=4$ supersymmetry, all of these analysis
are still based on the ${\cal{N}}=2$ superspace. Studies of the
next-to-leading correction in various branch are also important,
however, crossing quantum corrections among chiral multiplets and
hypermultiplets at four derivative order have not
been studied in detail because of the lack of simple ${\N}=2$
superspace formulation as far as we know.

On the other hand, splitting of quantum correction among $U(1)$
chiral multiplets and gauge neutral hypermultiplets at two
derivative order has been already observed in
\cite{decoupling, phase} since ${\N}=2$ supersymmetry implies
that the kinetic term of chiral multiplet is constrained to be
$\rm{\Kahler}$ geometry while that of hypermultiplet is to be 
hyper-$\rm{\Kahler}$ geometry. With the aid of
superspace Feynman rule, perturbative study of ${\N}=2$
non-holomorphic correction indicates
the K$\rm{\ddot{a}}$hler-like-geometry for four derivative 
terms involving only vector multiplets \cite{perturbative} and 
we then expect that this splitting structure is held
even at four derivative order. 

In this letter, we first show absence of crossing quantum corrections
among gauge neutral hypermultiplets and $U(1)$ vector multiplets on
${\N}=2$ non-holomorphic term at four derivative order.
We then revisit the non-renormalization theorem of four derivative
operator of gauge field for ${\N}=2$ SYM theory
\cite{non-renormalization theorem}
and clarify that it is true even though in mixed branch.

\section{Splitting of Quantum Corrections at Four Derivative
Order}
Let us begin with ${\N}=2$ SYM theory based on a group $G$
of rank $r$ which is spontaneously broken by the vacuum
expectation values of scalar fields either of chiral
multiplets or of hypermultiplets. In the following we focus
on such a mixed branch that is Abelian theory with $v(<r)$
Abelian chiral multiplets and $h$ gauge neutral
hypermultiplets. If the theory is 
asymptotically-free\footnote{In the scale-invariant case we
take the bare coupling to be small.},
regions of large vacuum expectation value of scalar fields
in chiral multiplet correspond to the weak coupling regime.
Taking all the vacuum expectation values large enough, all
massive fields become very heavy and almost decouple from
the theory. It is then reasonable that  the low energy 
``$Wilsonian$'' effective action is given
in terms of $v$ Abelian chiral multiplets and $h$ gauge
neutral hypermultiplet\cite{phase}. 
According to supersymmetric derivative expansion
\cite{hen}, four derivative operators would arise as
non-holomorphic corrections on ${\cal{N}}=2$ supersymmetric
effective action in the following form
\begin{equation}
\int d^4\theta d^4 \tilde{\theta} {\cal{H}}(\Psi^{a},
{\Psi^{a}}^{\dagger}; S^{i}, {S^{i}}^{\dagger}),
\end{equation}
where we denote gauge neutral ${\cal{N}}=2$ hypermultiplets by
$S^i$, $i=1,2,...,h$ and ${\N}=2$ Abelian chiral multiplets by
$\Psi^a$, $a=1,2,...,v$, respectively.
To analyze the structure of some scalar function ${\cal{H}}$,
it is convenient to replace ${\N}=2$ superfields in terms of
${\N}=1$ superfield\footnote{ We follow the conventions and
the reduction from ${\N}=2$ superfield to ${\N}=1$ superfield
of \cite{SW}.}and because of ${\N}=2$ supersymmetry, it is
enough to consider the following function \footnote{Due to
${\N}=2$ supersymmetry, we do not 
consider explicit dependence of ${\N}=1$ superfield strength 
$W_{\alpha}$ on ${\cal{H}}$ while for vector superfields $V$
we can show that the only possible dependence upon this function
is Fayet-Iliopoulos D-term by supersymmetric gauge invariance.}
\be
\int d^4 \theta d^4 \tilde{\theta} {\cal{H}}(\Phi^a,
{\Phi^a}^{\dagger}; Q^i, \tilde{Q^i}, {Q^i}^{\dagger},
\tilde{Q^i}^{\dagger}) ,
\ee
where we denote gauge neutral hypermultiplet by ${\N}=1$
chiral superfields $Q^i$, $\tilde{Q^i}^{\dagger}$, $i=1,2,...h$
and ${\N}=2$ Abelian chiral superfield by ${\N}=1$ superfields
$\Phi^a$, $W_{\alpha}^a$, $a=1,2,...,v$ respectively. 

Manipulating super-covariant derivatives, this includes a crossing
four derivative term, $\partial_{\phi} \partial_{\tilde{q}^{\dagger}}
{\cal{H}} \cdot \partial^2 \phi \partial^2 \tilde{q}^{\dagger}$
where $\phi$ and $\tilde{q}^{\dagger}$ are the scalar component of
chiral multiplet and that of hypermultiplet, respectively.
${\cal{N}}=1$ supersymmetry implies that
$\partial_{\phi} \partial_{\tilde{q}^{\dagger}} {\cal{H}} \cdot (-i
\bar{\zeta} \bar{\sigma}^{\mu} \partial_{\mu} \partial^2 \psi)$ must
be accompanied in order to cancel
${\cal{N}}=1$ supersymmetric variation, where we denote
hypermultiplet fermion and chiral multiplet fermion, associated
with ${\cal{N}}=1$ supersymmetry transformation, by $\bar\zeta$
and $\psi$, respectively. However, such a term is not allowed
since to cancel the other supersymmetric variation within 
${\cal{N}}=2$ supersymmetry, there must be a term involving four
derivatives, a gauge field and a scalar field, out of which no
Lorentz invariant combination can be formed. We then obtain
$\partial_{\phi} \partial_{\tilde{q}} {\cal{H}} = 0$.
This implies
\begin{equation}
{\cal{H}}(\Psi^a,{\Psi^a}^{\dagger}; S^i,{S^i}^{\dagger})
= 
{\cal{H}}_{V}(\Psi^a, {\Psi^a}^{\dagger}) + 
{\cal{H}}_{H}(S^i, {S^i}^{\dagger}),
\end{equation}
and we find the vacuum expectation value of hypermultiplet 
never appears in the ${\rm{\Kahler}}$ metric of chiral
multiplet even at four derivative order. Further
constraints will be given by thinking of coupling as a
background fields as we will see in the next section.

\section{Revisiting the Non-renormalization Theorem of Four
Derivative Operators in ${\cal{N}}=2$ finite SYM theory}

We would like to revisit the non-renormalization theorem
of four derivative operator of ${\N}=2$ finite SYM theory
in mixed branch as an immediate consequence of previous
section. Consider $SU(N)$ SYM theory, breaking to single
$U(1)$ case. With ${\N}=2$ superspace formulation, the
low energy effective action is
described by a $U(1)$ chiral multiplet $\Psi$ and some gauge
neutral massless hypermultiplets $S_i$. The kinetic term for
chiral superfield is given by the chiral integral
\be
\int d^2 \theta d^2 {\tilde{\theta}} \tau \Psi^2,
\ee
where $\tau$ is usual holomorphic gauge coupling \cite{SW}. 
Using the scale invariance and holomorphy, this term
remains quadratic after quantum corrections are included. It
is important to note that $\tau$ is chiral.
On the other hand, terms with four derivative arises from
\begin{equation}
\int d^4\theta d^4 \tilde{\theta} {\cal{H}}(\Psi,
{\Psi}^{\dagger}; S_i, {S_i}^{\dagger}) \label{eqn:1}.
\end{equation}
As we have shown in the previous section, no crossing term
among $U(1)$ chiral multiplet and gauge neutral hypermultiplets
is effectively generated, so four derivative terms of gauge
field arise from purely chiral part. We then 
restrict our attention to terms given by purely chiral multiplet
\be
\int d^4\theta d^4 \tilde{\theta} {\cal{H}}_{\rm{V}}(\Psi,
{\Psi}^{\dagger},\tau,\tau^{\dagger}).
\end{equation}
Both the scale-invariance and the $U(1)_{R}$
invariance provide a unique form 
\be
{\cal H} =c \ln \Psi \ln \Psi^{\dagger}
\label{eqn:a2},
\ee
with a constant $c$ \footnote{
Instanton calculation in \cite{instanton} further confirms
this result and determine the constant $c=1/(8\pi)^2$
\cite{const}.}.
To determine the $\tau$ dependence, we
promote $\tau$ as a background chiral superfield. This
expression spoils both the scale-invariance and the
$U(1)_{R}$ invariance unless it is independent of
$\tau$. Then we find that such a correction arises only 
from one-loop level.

\section{Remarks}
In this letter we have first presented absence of crossing
quantum corrections among $U(1)$ chiral multiplets and gauge
neutral hypermultiplets at four derivative order. From the
view point of symmetries of Lagrangian, we do not know how
they control the crossing quantum corrections among $U(1)$ 
chiral
multiplets and gauge neutral hypermultiplets but ${\N}=2$
supersymmetry controls them since a variation of Lagrangian
under supersymmetry is always total divergence and it is a
symmetry of action. The similar aspects can be seen in the
Chern-Simons Theory of three dimensional gauge theory \cite{CS}.

We then revisit the non-renormalization theorem given by
M. Dine and N. Seiberg. Our proof of the non-renormalization
theorem is given in terms of $Wilsonian$ effective action. The
$Wilsonian$ effective action is different from ordinary 1PI
effective action if interacting massless particles exist. As
reported in \cite{holomorphic-anomaly}, there is holomorphic
anomaly related to IR issues and resulting the violation of
non-renormalization theorem. Since they are absent in
$Wilsonian$ effective action, it is well-often favored in the
study of supersymmetric theories. However, as reported in
\cite{perturbative}, there are violating terms of special
K$\rm{\ddot{a}}$heler geometry even in a naive $Wilsonian$
effective action and authors of \cite{infrared} proposed the
$Wilsonian$ effective action with field dependent cut-off.
Further discussion about issues of reguralization is beyond
the scope of this letter.

The final remark is as follows: It is interesting to know
the non-renormalization theorem when the low energy effective
action is described by more than two $U(1)$ chiral superfields.
In our proof of non-renormalization theorem in mixed branch
we have simply restricted to $SU(N)$ breaking to single $U(1)$ case.
Since renormalization is UV nature, it is independent
of IR structure and there might be non-renormalization theorem
even if more than two $U(1)$ gauge symmetries survive.
Unfortunately, we do not know how to obtain the exact
solution by utilizing symmetries of Lagrangian. For example,
Such a term
\begin{equation}
f(\tau, \tau^{\dagger}) \frac{\Psi^a}{\Psi^b}
\frac{\bar{\Psi}^c}{\bar{\Psi}^d}, \label{eq:1}
\end{equation}
do not conflict with both the scale-invariance and the
$U(1)_{R}$ invariance.  In perturbative study of ${\N}=2$
SYM theory\cite{extension}, the terms like (\ref{eq:1}) is absent
but how it goes non-perturbatively? Direct test by instanton
calculation is of great interest.
%
%
%
%
\subsection*{Acknowledgement}

We would like to thank for M. Nitta and M. Sakamoto
for useful comments and fruitful discussions.

\end{document}